\documentstyle[preprint,eqsecnum,aps]{revtex}

\author{Xun Xue$^{1,3}$ and\ 
A. Ferraz$^{1,2}$\\ 
$^1$International Centre of Condensed Matter Physics\\
Universidade de Brasilia, Caixa Postal 04667\\
70910-900 Brasilia - DF - Brazil\\
$^2$Yukawa Institute for Theoretical Physics\\
Kyoto University, Kyoto 606, Japan\\
$^3$Advanced Research Centre\\ 
Zhongshan(Sun Yat-sen) University, Guangzhou, China}
\title{\bf SUPERSYMMETRY AND ELECTRON-HOLE  EXCITATIONS IN SEMICONDUCTORS
AT FINITE TEMPERATURE}
\input tcilatex

\QQQ{Language}{
American English
}

\begin{document}

\maketitle
\begin{abstract}
The fermionic and bosonic electron-hole low lying excitations in a
semiconductor are analyzed at finite temperature in a unified way 
following Nambu's
quasi-supersymmetric approach for the BCS model of superconductivity. 
The effective lagrangian for the fermionic
modes and for the bosonic low lying 
collective excitations in the semiconductor is no longer supersymmetric
in a conventional  
finite temperature treatment. However the bosonic excitations don't 
couple directly to the heat bath and as a result, quasisupersymmetry 
is restored to the effective lagrangian
when a redefinition of the coupling constant associated with the 
collective excitations is performed. Our result shows that
although the mass and coupling parameters are now temperature dependent, 
the fermion and 
boson excited states pair together and can still be transmuted into one 
another. \

\end{abstract}


\section{Introduction}

It can be shown that a model with a BCS type of spontaneously symmetry 
breaking possesses a quasi-supersymmetry if there exists an appropriate mass 
ratio 
relation for the bosonic and fermionic excitations 
of the physical system \cite{b1,b2,b3,b4}. Nambu investigated the BCS model 
of superconductivity in which the masses of 
the three low energy excitations, the phase 
($\pi$ or Goldstone) collective mode, the fermionic excitation(the 
quasiparticle) and the amplitude($\sigma$ or Higgs) mode
are in the ratios $0:1:2$.
He showed that the static part of the BCS model can be expressed as an
anticommutation of the fermionic composite charges $Q$ and $Q^{+}$. The 
$Q$'s are not nilpotent but the equal masses of a pair of fermionic 
excitations
and their corresponding bosonic mode is a clear sign of the built-in
quasi-supersymmetry in this model.
In our previous work \cite{b5}, we invoke the pairing approximation,
to make physically equivalent the fermion operators, hole annihilation 
$\psi _h$
and electron creation $\psi _e^{+}$, when they act in an appropriate 
excited state
function in the semiconductor. When this equivalence relation holds, it 
can be shown that the static part of the corresponding
effective hamiltonian is truly supersymmetric at $T=0$. The boson and
fermion excitations can be mapped into one another and the model is
identical to $1+1$ supersymmetric quantum mechanics($SUSY$).

In a pure semiconductor, the low lying
electronic excitation, as is well known, is a bound electron-hole pair 
state called exciton \cite{b6,b7,b8,b9}.
The excitons can be well characterized by their
radii. If the exciton radius is of the order of the lattice constant the
pair is tightly bound and the exciton state is known as Frenkel exciton. In
contrast, in a semiconductor with a large dielectric constant this radius
extends over many lattice cells, and as a result the pair is only weakly
bound, and known as Wannier exciton. We confine ourselves to this case
only. 

In our previous work the electron-hole pair, the single-particle states 
and the
phase mode, which is transmuted into a plasmon mode in the presence of
Coulomb forces, are described in an unified manner. For a direct gap
semiconductor, we show that, similarly to superconductivity, the single
fermion and boson masses also define a $0:1:2$ ratio at zero-temperature.

However, it is known that even if supersymmetry exists at zero-temperature 
it will be spontaneously broken at finite temperature
due to the fact that fermions and bosons are clearly distinguishable 
from each other when 
they couple to a heat bath. As a result their effective masses therefore no 
longer maintain the same 
value since they obey different statistics \cite{b18}. 
So, at a first glance, the model which we investigated in 
our previous work is not, strictly speaking, supersymmetric at finite 
temperatures.

Considering that our effective action model describes low energy 
fermionic and bosonic
excitations in the semiconductor, the finite 
temperature$^{\prime}$s 
main effect is in general terms to reduce the magnitudes of both the low 
lying excitation energies and the 
energy gap. The basic physical picture involving 
the electron, the hole, the exciton and the plasmon excitations remains 
practically unchanged at $T\ne 0$. It is reasonable to expect 
that the mass relation and the supersymmmetry will be maintained in a 
properly redefined effective 
action for  the physically relevant low lying fermionic and bosonic 
elementary excitations at finite T.

In the present work, we first derive from first principles, the finite 
temperature form of the effective action 
and show that the mass relation between the existing excitations is 
lost. If this were the case supersymmetry would be spontaneously broken at 
any finite temperature in the semiconductor. 
This generalized model however does not reveal the real physical 
properties of 
the low energy lying excitations in the semiconductor. We are 
therefore led to 
construct a new effective action for the low energy lying excitations of the 
semiconductor at 
finite temperatures. When we do this, the model is shown to maintain both 
its supersymmetry 
and the mass relations as proposed earlier by Nambu.

\section{EFFECTIVE ACTION FOR ELECTRON-HOLE AND THEIR COLLECTIVE EXCITATIONS}

Let us review briefly the derivation of the effective action at $T=0$.
Consider for simplicity a two-band direct semiconductor with parabolic
conduction and valence bands. At $T=0$, the valence band is completely full
and the conduction band is totally empty. If we excite the semiconductor by
laser pumping or by any other means, producing interband transitions between
these two bands, the single-particle picture of the resulting low-lying
state is basically given in terms of the promotion of the electron to the
conduction band and the creation of a hole in the valence band. In reality
this single-particle picture is known to be incomplete since electron and
hole interact strongly with each other and the resulting pair of particles
is associated primarily with a $1s$ type level inside the energy gap.

Let us then consider the model hamiltonian density ${\cal {H}}_f$ given by

\begin{equation}
\label{e21}{\cal H}_f=\psi _e^{+}\epsilon \psi _e+\psi _h^{+}\epsilon \psi
_h-\lambda \psi _e^{+}\psi _h^{+}\psi _h\psi _e\text{ ,} 
\end{equation}
which describes the low-lying electron-hole excitations in the
semiconductor. Here, for convenience, we take equal electron and hole
effective masses $m_e=m_h=m$ and, in k-space, $\epsilon =\frac{E_g}2+\frac{%
k^2}{2m}$, where $E_g$ is the energy gap and $\lambda $ is an effective
four-fermion coupling constant.

Following standard procedure we define next the path integral $Z$ for these
electron and hole states:

\begin{equation}
\label{e22}Z=\int {\cal D}\psi _e^{+}{\cal D}\psi _e{\cal D}\psi _h^{+}{\cal %
D}\psi _h\exp i\int {\cal L}_f 
\end{equation}
where

\begin{equation}
\label{e23}{\cal L}{_f}=\psi _e^{+}(i{\partial _t}-\epsilon )\psi _e+\psi
_h^{+}(i{\partial _t}-\epsilon )\psi _h+\lambda \psi _e^{+}\psi _h^{+}\psi
_e\psi _h 
\end{equation}
is the lagrangian density corresponding to the model hamiltonian 
$(\ref{e21})$. In 
order to evaluate the path integral $Z$ it is useful to rewrite the
quartic interaction in terms of effective quadratic terms. For this we use
the Hubbard-Stratonovich transformation \cite{b10,b11} for the bosonic 
auxiliary complex fields $\phi ,\phi ^{*}$:

$$
\int {\cal D}\phi {\cal D}\phi ^{*}\exp -i\int \left[m_0^2 \phi ^{*}\phi -m_0\sqrt{%
\lambda }\psi _e^{+}\psi _h^{+}\phi -m_0\sqrt{\lambda }\phi ^{*}\psi _h\psi
_e\right] 
$$

\begin{equation}
\label{e24}
=const\exp \left( i\lambda \int \psi _e^{+}\psi _h^{+}\psi _h\psi _e\right) 
\text{.} 
\end{equation}
When we replace this identity in the path integral, $Z$ becomes

\begin{equation}
\label{e25}Z=\int {\cal D}\psi _e^{+}{\cal D}\psi _e{\cal D}\psi _h^{+}{\cal %
D}\psi _h{\cal D}\phi ^{*}{\cal D}\phi \exp i\int {\cal L}_{\psi -\phi } 
\end{equation}
where

\begin{equation}
\label{e26}{\cal L}_{\psi -\phi }=\psi _e^{+}(i{\partial _t}-\epsilon )\psi
_e+\psi _h^{+}(i{\partial _t}-\epsilon )\psi _h+m_0{\sqrt{\lambda }}\psi
_e^{+}\psi _h^{+}\phi +m_0{\sqrt{\lambda }\phi ^{*}}\psi _h\psi _e-m_0^2\phi
^{*}\phi 
\end{equation}
Using the Nambu matrix notation for the fermion field $\Psi$,

$$
\Psi =\left( 
\begin{array}{c}
\psi _e \\ 
\psi _h^{+} 
\end{array}
\right) \text{,}
$$
the lagrangian ${\cal L}_{\psi -\phi }$ becomes

\begin{equation}
\label{e27}{\cal L}_{\psi -\phi }={\Psi ^{+}}\left[ (i{\partial _t}-\epsilon
\tau _3)+m_0{\sqrt{\lambda }}(\phi {\tau _{+}}+\phi ^{*}{\tau _{-}})\right]
\Psi -{m_0}^2\phi ^{*}\phi +const 
\end{equation}
where ${\tau _{\pm }}=\frac 12({\tau _1}\pm i{\tau _2})$ with the ${\tau _i}%
^{\prime }s$ $(i=1,2,3)$ being the Pauli matrices. The fermionic lagrangian
is in this way transformed into an effective fermion-boson model with a
Yukawa-type interaction which is bilinear in the fermion fields.

Substituting this lagrangian in the path integral $Z$ and integrating it
over the fermion fields, it then follows that

\begin{equation}
\label{e28}Z=\int {\cal D}\phi {\cal D}\phi ^{*}\exp -iW(\phi ^{*},\phi ) 
\end{equation}
where $W$ is now the effective `action' for the bose fields $\phi $ and $%
\phi ^{*}$ given by

\begin{equation}
\label{e29}W(\phi ^{*},\phi )=\int m_0^2\phi ^{*}\phi +iTrln\left[ i\partial {_t}%
-\epsilon \tau _3+m_0{\sqrt{\lambda }}(\phi {\tau _{+}}+\phi ^{*}{\tau _{-}}%
)\right] 
\end{equation}

To determine the low energy and long wavelength dynamics of the bosonic 
fields $\phi$ 
and $\phi^*$, we will make a gradient expansion in the effective
action W($\phi^{*}$, $\phi$) to derive their kinetic energy term  
\cite{b10,b11}. Let us decompose the complex scalar field in the form

\begin{equation}
\label{e210}\phi =\frac 1{\sqrt{2}}(\phi _1+i\phi _2),\;\;\phi ^{*}=\frac 1{
\sqrt{2}}(\phi _1-i\phi _2)\text{ ,} 
\end{equation}
and define the interaction matrix operator $M$ as

\begin{equation}
\label{e211}M=m_0\sqrt{\lambda }(\phi \tau _{+}+\phi ^{*}\tau _{-})=m_0\sqrt{\lambda }\;%
{\underline{\phi }}.{\underline{\tau }}\text{ ,} 
\end{equation}
where $\underline{\phi }$ is the two-component real vector field

\begin{equation}
\label{e212}\underline{\phi }=\frac 1{\sqrt{2}}\left( \phi _1,-\phi _2,0\right) \text{ .}
\end{equation}

It then follows that

\begin{equation}
\label{e213}W(\phi )=iTrln[k_0-\epsilon _k\tau _3]-i\sum_{n=1}\frac{(i)^{n}}%
nTr(G_0 M)^n+\int m_0^2\underline{\phi }.\underline{\phi }, 
\end{equation}

where

\begin{equation}
\label{e214}G_0\left( k\right) =i\text{ }\frac{k_0+\epsilon _{{\bf k}%
}\tau _3}{k_0^2-\epsilon_{%
{\bf k}}^2+i\delta } 
\end{equation}
is the fermionic Green's function.

Up to the quartic term in ${\bf \phi }$, our effective action
can be written as

\begin{equation}
\label{e215}W(\underline{\phi })=\int\limits_q
m_0^2\phi ^a(q)K_{ab}(q)\phi ^b(q)
+\frac 14\Lambda (0)_{abcd}%
\phi^a \phi^b \phi^c \phi^d 
\end{equation}
where we use a notation $\underline{\phi}=(\phi^1,\phi^2,0)$,

\begin{equation}
\label{e216}K_{ab}(q)=\delta _{ab}+\frac \lambda 2\pi
_{ab}(q)\text{,}
\end{equation}
with

\begin{equation}
\label{e217}\pi _{ab}(q)=i\int\limits_kTr[G_0(k+q)\tau
_aG_0(k)\tau _b] \text{,}
\end{equation}
represented by the bubble diagram in Fig. 1, and

\begin{equation}
\label{e218}\Lambda (0)_{abcd}={-im_0^4\lambda ^2}\int_kTr[G_0(k)\tau
_aG_0(k)\tau _bG_0(k)\tau _cG_0(k)\tau _d] 
\end{equation}
as is shown in Fig. 2. It is easy to show that the linear and cubic 
term in $\phi$ in the expansion of $W(\phi)$ now give zero contributions.

We then have that 
\begin{equation}
\label{e219}
\Lambda (0)=-2im_0^4\lambda^2\int\limits_k\frac{1}{(k_0^2-%
\epsilon_{\bf k}^2)^2}
\end{equation}

and 

\begin{equation}
\label{e220}\pi_{ab} (q)=\pi_{ab} ^{(1)}(q)+\pi_{ab} ^{(2)}(q)
\end{equation}

where

\begin{equation}
\label{e221}\pi_{ab} ^{(1)}(q)=-2i\int\limits_k\frac{k_0(k_0+q_0)-%
\epsilon {\bf _k}\epsilon _{{\bf k}+{\bf q}}}{%
[(k_0+q_0)^2-\epsilon_{{\bf k+q}}^2][k_0^2-\epsilon_{{\bf k}}^2]}%
\delta_{ab}=\pi^{0}(q)\delta_{ab} 
\end{equation}
and

\begin{equation}
\label{e222}\pi_{ab} ^{(2)}(q)=-2 \int\limits_k\frac
{(k_0+q_0)\epsilon_{{\bf k}}-k_0\epsilon_{{\bf k}+{\bf q}}}{%
[(k_0+q_0)^2-\epsilon_{{\bf k}+{\bf q}}^2][k_0^2-\epsilon_{{\bf k}}]}%
\epsilon_{ab}\;\;\;.
\end{equation}

It is sufficient to expand $\pi ^{(0)}$ 
and $\pi_{ab}^{(2)}$ around $q_0=0$ and ${\bf q}=0$
to derive the kinetic and the time derivative
terms for the bosonic field. 
Making the expansion of $\pi ^{(0)}$
in terms of the gradients of $\underline {\phi}$, we get

\begin{equation}
\label{e223}K^{(0)}(0)=(1+\frac {\lambda}{2}\pi^{(0)}(0))
\end{equation}
and

\begin{equation}
\label{e224}K^{(0)}(q)=Zq_0^2+I{\bf q^2}+\cdots 
\end{equation}
where

\begin{equation}
\label{e225}\pi^{(0)}(0)=-2i\int\limits_k\frac{1}{k_0^2-\epsilon_{{\bf k}}%
^2+i\delta}\;\;\;,
\end{equation}

\begin{equation}
\label{e226}Z=-i\lambda \int_k\frac{k_0^2+\epsilon_{{\bf k}}^2}%
{(k_0^2-\epsilon_{{\bf k}%
}^2)^3}\;\;\;, 
\end{equation}
and

\begin{equation}
\label{e227}I=\frac{-i\lambda }{2m}\int_k\frac{\epsilon _{{\bf k}}}{%
(k_0^2-\epsilon_{{\bf k}}^2)^2}+\frac{-i\lambda }{m^2}\left[ 2\int_k%
\frac{\epsilon
_{{\bf k}}^2{\bf k}^2}{(k_0^2-\epsilon_{{\bf k}}^2)^3}+%
\int_k\frac{{\bf k}^2}{%
(k_0^2-\epsilon_{{\bf k}}^2)^2}\right] 
\end{equation}
with $\pi_{ab}^{(2)}$ giving no contribution.

Evaluating the integrals above at zero temperature with the Fermi
energy $\epsilon _F$ as a cut-off parameter , for one-electron states near
the bottom of conduction band (i.e. for $\frac 12E_g>>\epsilon _F%
$) we find

\begin{equation}
\label{e228}K^{(0)}(0)=\left[1-\frac{\lambda n}{E_g}\left(1-\frac35%
\frac{\epsilon_F}{E_g/2}\right)\right]
\end{equation}
where $n=(2m\epsilon_F)^{3/2}/6\pi^2$ is the free particle density in 
conduction band,
\begin{equation}
\label{e229}
Z=\frac{-\lambda }{6\pi ^2}\frac{(2m\epsilon _F)^{3/2}}{E_g^3}\left( 1-\frac
95\frac{\epsilon _F}{E_g/2}\right)\;\;\;, 
\end{equation}

\begin{equation}
\label{e230}
I=\frac \lambda {6\pi ^2}(2m)^{1/2}\frac{\epsilon _F^{3/2}}{E_g^2}\left( 1- 
\frac{12}5\frac{\epsilon _F}{E_g/2}\right)
\end{equation}

and

\begin{equation}
\label{e231}
\Lambda (0)=-4\lambda m_0^4Z\;\;\;.  
\end{equation}

Rescaling the field and coupling constants as

\begin{equation}
\label{e232}
m_0\underline {\phi} =(-Z)^{-1/2} \underline{\varphi} \text{,} 
\end{equation}

\begin{equation}
\label{e233}
\lambda (-Z)^{-1}=f^2 
\end{equation}
and

\begin{equation}
\label{e234}
\frac{K^{(0)}(0)}{Z}=m^{\prime 2}\text{ ,} 
\end{equation}
the effective lagrangian for the bosonic modes in the semiconductor apart
from an unimportant constant is

\begin{equation}
\label{e235}{\cal L_{\underline{\varphi}}}=%
(\partial _t{\underline{\varphi}} )^2-\alpha ^2({\bf% 
{\nabla }}{\underline{\varphi}})\cdot({\bf {\nabla }}%
{\underline{\varphi}} )-
f^2\left( {\underline{\varphi}} ^2-\frac{m^{^{\prime }2}}{2f^2}\right) ^2 
\end{equation}
which is of a Ginzburg-Landau form,
with $\alpha ^2=\frac 1m(\frac 12E_g-\frac 35\epsilon _F)$.

It should be noted that the contribution from $K^{(0)}(q=0)$ 
gives a virtual mass constant
and thus there is no spontaneous symmetry breaking unless the coupling
constant
$\lambda$ exceeds a critical value 
\begin{equation}
\label{e236}\lambda_c=\frac{E_g}n\left(1+\frac35%
\frac{\epsilon_F}{E_g/2}\right)\text{.}
\end{equation}

Making the substitution of rescaled field 
$\underline{\varphi}$ with the previous notation $\underline{\phi}$
and using $\underline{\pi}$ as the canonical conjugate to $\underline{\phi}$,
the effective hamiltonian density ${\cal H}_{\psi
-\phi }$ for the fermionic and bosonic excitations in the semiconductor is

\begin{equation}
\label{e237}{\cal H}_{\psi -\phi }=\psi ^{+}(\epsilon \tau _3+%
f\underline{\phi}\cdot \underline{\tau} )\psi +\underline{\pi} ^2+\alpha ^2%
({\bf \nabla} {\underline{\phi}} )\cdot ({\bf \nabla }{\underline{\phi}} )+%
f^2({\underline{\phi}} ^2-\left( \frac{%
m^{^{\prime }}}{\sqrt{2}f}\right) ^2)^2\text{,} 
\end{equation}
or, equivalently, in terms of $\phi$ and $\phi^*$, we get

\begin{equation}
\label{e238}{\cal H}_{\psi -\phi }=\psi ^{+}(\epsilon \tau _3+%
f\phi \tau_+ +f\phi^* \tau_-)\psi +\pi^* \pi+\alpha ^2%
({\bf \nabla} \phi^* )\cdot ({\bf \nabla} \phi )+%
f^2(\phi^* \phi-\left( \frac{%
m^{^{\prime }}}{\sqrt{2}f}\right) ^2)^2\text{.} 
\end{equation}

This Hamiltonian can be mapped into 
a supersymmetric quantum mechanical model if we follow the procedure used 
in our earlier work \cite{b5}.

\section{EFFECTIVE HAMILTONIAN FOR ELECTRON-HOLE AND THEIR COLLECTIVE 
MODE EXCITATIONS AT FINITE TEMPERATURE}

Let us now approach the finite temperature case. For this, we first 
derive the temperature dependent effective action for the bosonic field 
following out our procedure in Sec.II. Finite temperature effects can be 
obtained by making the field 
configurations periodic or anti-periodic in time for bosons or femions 
respectively, with a period $\beta=1/T$(with the Boltzmann constant 
being equal to one), in the imaginary time formalism. In momentum space 
what is 
needed is the replacement of $\int dk_0/(2\pi)$ by $i/\beta \sum_n $ where 
$k_0=\frac{2\pi i}{\beta} (n+\frac{1}{2})$, for fermions, and 
$k_0=\frac{2\pi i}{\beta} n$, for bosons \cite{b12}. Making these 
substitutions we can easily get the finite temperature
form of the quantities we are interested in, namely
\begin{equation}
\label{e31}\pi^{(0)}_T(0)=-\frac{(2m\epsilon_F)^{3/2}}{6\pi^2(E_g/2)}%
\left[ \left(1-\frac35%
\frac{\epsilon_F}{E_g/2}\right)tanh(\frac{E_g}{4T})+\frac35%
\frac{\epsilon_F}{2T}\frac{1}{cosh^2\left(\frac{E_g}{4T}\right)}\right]%
\text{,}
\end{equation}

\begin{equation}
\label{e32}K^{(0)}_T(0)=1-\frac{\lambda n}{E_g}\left[\left(1-\frac35%
\frac{\epsilon_F}{E_g/2}\right)tanh(\frac{E_g}{4T})+\frac35%
\frac{\epsilon_F}{2T}\frac{1}{cosh^2\left(\frac{E_g}{4T}\right)}\right]%
\text{,}
\end{equation}
where $n=(2m\epsilon_F)^{3/2}/6\pi^2$,

$$
Z(T)=\frac{-\lambda }{6\pi ^2}\frac{(2m\epsilon _F)^{3/2}}{E_g^3}%
\left[\left( 1-\frac95\frac{\epsilon _F}{E_g/2}\right)\left( tanh\left(\frac{E_g}{4T}\right)%
-\frac{E_g}{2T}\frac{1}{cosh^2\left(\frac{E_g}{4T}\right)}\right)\right.
$$
\begin{equation}
\label{e33}\left.-\left(1-\frac35 \frac{\epsilon_F}{E_g/2}\right)%
\frac{(E_g/2)^2}{T^2}\frac{tanh(E_g/4T)}%
{cosh^2\left(E_g/4T\right)}+\frac35\frac{\epsilon_F(E_g/2)^2}{4T^3}%
\frac{1-sinh^2(E_g/4T)}{cosh^4(E_g/4T)}\right]\text{,}
\end{equation}

$$
I(T)=\frac \lambda {6\pi ^2}(2m)^{1/2}\frac{\epsilon _F^{3/2}}{E_g^2}%
\left[\left( 1-\frac{12}5\frac{\epsilon _F}{E_g/2}\right)tanh(E_g/4T)\right.
$$
\begin{equation}
\label{e34}\left.-\frac{E_g}{4T}\left(1-\frac{36}{5} \frac{\epsilon_F}{E_g/2}\right)%
\frac{1}{cosh^2\left(E_g/4T\right)}%
+\frac35 \frac{\epsilon_F(E_g/2)}{T^2}\frac{tanh(E_g/4T)}%
{cosh^2\left(E_g/4T\right)}\right]\text{,}
\end{equation}
and finally
\begin{equation}
\label{e35}\Lambda_T (0)=-4\lambda m_0^4\left(Z(T)+\frac{3\lambda n \epsilon _F}%
{16\cdot 5E_gT^3}%
\frac{1-sinh^2(E_g/4T)}{cosh^4(E_g/4T)}\right)\text{.}
\end{equation}
Now if we rescale the field using
\begin{equation}
\label{e36}m_0\underline {\phi} =(-Z(T))^{-1/2} \underline{\varphi} \text{,} 
\end{equation}

\begin{equation}
\label{e37}m_0^2\lambda (-Z(T))^{-1}=f_T^2 \text{,}
\end{equation}

\begin{equation}
\label{e38}\frac{\Lambda_T (0)}{4m_0^4} (-Z(T))^{-2}=f(T)^2 
\end{equation}
and

\begin{equation}
\label{e39}\frac{K^{(0)}_T(0)}{Z(T)}=m^{\prime 2}_T\text{ ,} 
\end{equation}
the resulting effective Hamiltonian for the bosonic and fermionic 
excitations becomes 
$$
{\cal H}_{\psi -\phi }=\psi ^{+}(\epsilon(T) \tau _3+%
f_T\underline{\phi}\cdot \underline{\tau} )\psi +\underline{\pi} \cdot\underline{\pi}
$$
\begin{equation}
\label{e332}+\alpha_T ^2%
({\bf \nabla} {\underline{\phi}} )\cdot ({\bf \nabla }{\underline{\phi}} )+%
f(T)^2\left({\underline{\phi}} ^2-\left( \frac{%
m^{^{\prime }}_T}{\sqrt{2}f(T)}\right) ^2\right)^2\text{,}
\end{equation}

or in our $\phi$ and $\phi^*$ notation from Section $\bf {II}$,
$$
{\cal H}_{\psi -\phi }=\psi ^{+}(\epsilon(T) \tau _3+%
f_T\phi \tau_+ +f_T\phi^* \tau_-)\psi +\pi^* \pi
$$
\begin{equation}
\label{e333}+\alpha_T ^2%
({\bf \nabla} \phi^* )\cdot ({\bf \nabla} \phi )+%
f(T)^2\left(\phi^* \phi-\left( \frac{%
m^{^{\prime }}_T}{\sqrt{2}f(T)}\right) ^2\right)^2\text{,} 
\end{equation}
where $\alpha_T ^2=-\frac{I(T)}{Z(T)}$.
In this present form the model can no longer be mapped into $1+1$ 
supersymmetric 
quantum model since the key feature $f_T=f(T)$ does not hold anymore at any 
finite temperature. This 
is crucial for a model to satisfy the requirements for 
quasisupersymmetry, 
and hence Nambu$^{\prime}$s argument on the realization on the 
quasisupersymmetry is no longer applicable. This result is not 
unexpected since in our derivation of the model
both fermion and boson are coupled directly to the heat bath. 
Due to 
their different statistics they necessarily acquire different 
effective masses. The supersymmetry of zero temperature model is therefore
spontaneously broken by finite temperature effects if we assume 
that both single particle and electron-hole pair states couple 
independently to the heat bath.

However it is not physically meaningful to do this. As in a superconductor 
in which the low-lying excitations of the BCS ground state are 
associated with the breaking of Cooper pairs and not with higher energy 
paired states, here, near the semiconductor-metal transition, we are in 
a somewhat similar situation. It is precisely the breaking of 
electron-hole pair which produces new excited states which are 
closer in energy to the supersymmetric reference state. The main reason 
for this is that the original single-particle states are the real 
fundamental entities in the electron-hole pair. Although the paired state 
can be described by a boson field the bosonic high energy states are not 
in reality
accessible to the physical system. In other words if we couple both the 
fermion and boson states directly to the heat bath, the model needs to be 
corrected further since the higher energy bosonic states have to be 
excluded from the Hilbert space associated with our effective hamiltonian. 
One alternative way to achieve this is to consider that the bosonic field 
only couples indirectly to the heat bath through 
the fermionic excitation which feels the presence of a non-zero 
temperature. If we couple the fermions to the heat bath in the first place 
we no longer need to do the same with the boson excitations. The bosons 
will automatically feel the finite temperature effects through the 
redefined coupling and mass parameters. Let us go back to the original 
fermionic model at $T=0$:

\begin{equation}
\label{e312}{\cal H}_f=\psi _e^{+}\epsilon \psi _e+\psi _h^{+}\epsilon \psi
_h-\lambda \psi _e^{+}\psi _h^{+}\psi _h\psi _e\text{ .} 
\end{equation}
It describes the low-lying electron-hole excitations in the semiconductor
with $\epsilon=\frac{E_g}2+\frac{k^2}{2m}$. It is 
sufficient for us to study what happens to this simple 
Hamiltonian density at finite temperature.
We do not need to care about the precise form of $\epsilon$ at finite 
$T$.
Let us denote it by $\epsilon(T)$. In the imaginary time formalism, the 
finite
temperature effect will come as a result of the loop corrections 
performed at non-zero temperature. For simplicity, we will only consider
the one-loop correction to the four-fermion coupling constant $\lambda$, 
as is shown in Fig. 3.

The resulting $\lambda_{1-loop}$ is given in the following form with simple 
derivation \begin{equation}
\label{e313}\lambda_{1-loop}=\lambda-2\lambda^2 \int_k%
\left[\frac1{(k_0^2-\epsilon_{{\bf k}}^2)}%
+\frac{2\epsilon_{{\bf k}}}{(k_0^2-\epsilon_{{\bf k}}^2)^2}\right]\;\;\;, 
\end{equation}

using the fermionic propagator 
$$G_0\left( k\right) =i\text{ }\frac{k_0+\epsilon _{{\bf k}%
}\tau _3}{k_0^2-\epsilon_{{\bf k}}^2+i\delta }\text{.}$$

Performing finite temperature calculation to $\lambda_{1-loop}$ in the 
imaginary time formalism, we get 
\begin{equation}
\label{e314}\lambda_T=\lambda+\frac{\lambda^2}{2T^2} \int_{{\bf k}}%
\frac1{cosh^2(\epsilon_{{\bf k}}/2T)}\;\;\; \text{.}
\end{equation}
Evaluating the integral over ${\bf k}$, the finite temperature effective 
four fermion coupling constant is
\begin{equation}
\label{e315}\lambda_T=\lambda+\frac{\lambda^2n_T}{2T^2}\left[%
\frac1{cosh^2(E_g/4T)}-\frac35\frac{\epsilon_F}T%
\frac{tanh(E_g/4T)}{cosh^2(E_g/4T)}\right]\;\;\; 
\end{equation}
where $n_T=(2m\epsilon_F)^{3/2}/6\pi^2$ and $\epsilon_F$ and $E_g$ take 
their corresponding values at temperature T.

At this stage we can repeat all the computation of Sec.II to derive the 
effective
action for the fermionic and bosonic fields with a temperature dependent 
coupling constant. It then follows that
\begin{equation}
\label{e316}K^{(0)}_T(0)=\left[1-\frac{\lambda_T n_T}{E_g}\left(1-\frac35%
\frac{\epsilon_F}{E_g/2}\right)\right]\text{,}
\end{equation}

\begin{equation}
\label{e317}
Z_T=\frac{-\lambda_T }{6\pi ^2}\frac{(2m\epsilon _F)^{3/2}}{E_g^3}\left( 1-\frac
95\frac{\epsilon_F}{E_g/2}\right)\;\;\;, 
\end{equation}

\begin{equation}
\label{e318}
I_T=\frac {\lambda_T} {6\pi ^2}(2m)^{1/2}\frac{\epsilon _F^{3/2}}{E_g^2}\left( 1- 
\frac{12}5\frac{\epsilon _F}{E_g/2}\right)
\end{equation}

and

\begin{equation}
\label{e319}
\Lambda_T (0)=-4\lambda_T m_0^4Z_T\;\;\;.  
\end{equation}

Rescaling the field and coupling constants in the same way as before, we get

\begin{equation}
\label{e320}
m_0\underline {\phi} =(-Z_T)^{-1/2} \underline{\varphi} \text{,} 
\end{equation}

\begin{equation}
\label{e321}
\lambda_T (-Z_T)^{-1}=f_T^2 
\end{equation}
and

\begin{equation}
\label{e322}
\frac{K^{(0)}_T(0)}{Z_T}=m^{\prime 2}_T\text{ .} 
\end{equation}
The effective lagrangian for the bosonic modes in the semiconductor apart
from an unimportant constant will again have a Ginzburg-Landau form which  
differs only
from its zero $T$ partner by the fact that the masses and coupling 
parameters are now temperature dependent.

The final effective Hamiltonian density ${\cal H}_{\psi
-\phi }$ for the fermionic and bosonic excitations in the semiconductor 
at $T\ne 0$ is then
$$ 
{\cal H}_{\psi -\phi }=\psi ^{+}(\epsilon_T \tau _3+%
f_T\underline{\phi}\cdot \underline{\tau} )\psi +\underline{\pi} \cdot\underline{\pi}
$$
\begin{equation}
\label{e323}+\alpha_T ^2%
({\bf \nabla} {\underline{\phi}} )\cdot ({\bf \nabla }{\underline{\phi}} )+%
f_T^2\left({\underline{\phi}} ^2-\left( \frac{%
m^{^{\prime }}_T}{\sqrt{2}f_T}\right) ^2\right)^2\text{,}
\end{equation}
or, equivalently
$$
{\cal H}_{\psi -\phi }=\psi ^{+}(\epsilon_T \tau _3+%
f_T\phi \tau_+ +f_T\phi^* \tau_-)\psi +\pi^* \pi
$$
\begin{equation}
\label{e324}+\alpha_T ^2%
({\bf \nabla} \phi^* )\cdot ({\bf \nabla} \phi )+%
f_T^2\left(\phi^* \phi-\left( \frac{%
m^{^{\prime }}_T}{\sqrt{2}f_T}\right) ^2\right)^2\text{,} 
\end{equation}

where we have proceeded with the substitution of rescaled field 
$\underline{\varphi}$ by $\underline{\phi}$
with $\underline{\pi}$ being its canonical conjugate
and with $\alpha_T ^2=\frac 1m(\frac 12E_g-\frac 35\epsilon _F)$.

It should be noted that there is no spontaneous symmetry breaking
unless the mass term contributed by $K^{(0)}_T(q=0)$ is negative. 
This means that the  coupling constant
$\lambda_T$ should exceed a critical value 
\begin{equation}
\label{e325}\Lambda_c=%
\lambda_c+\frac{\lambda_c^2n_T}{2T^2}\left[%
\frac1{cosh^2(E_g/4T)}-\frac35\frac{\epsilon_F}T%
\frac{tanh(E_g/4T)}{cosh^2(E_g/4T)}\right]%
=\frac{E_g}{n_T}\left(1+\frac35%
\frac{\epsilon_F}{E_g/2}\right)\text{,}
\end{equation}
in order for this to occur.

If we now repeat 
the discussion employed in our $T=0$ treatment of the low-lying 
excitation and the many-body excited state vector near the 
semiconductor-metal transition\cite{b5} this supersymmetry can be made exact 
even in the presence of a heat bath.
>From the discussion in the last two sections and
from Nambu$^{\prime}$s argument on the BCS model of
superconductivity, it will soon become clear that we have derived a 
quasi-supersymmetric model for the low energy fermionic and bosonic 
excitations in a semiconductor at finite temperature T.

\section{State Vector and Low-Lying Excitations in a Semiconductor}

It is well known that the low-lying excitations in a semiconductor are
mostly associated with exciton states. 
Let $|\Phi _0>$ represent
the ground-state of the semiconductor. Then

\begin{equation}
\left| \Phi _0\right\rangle =\left| 0\right\rangle \otimes \prod\limits_{k 
\text{ }(k_0\leq -\frac 12E_g)}\psi _e^{+}(k_0,{\bf k)}\left| 0\right\rangle
\equiv \prod\limits_{k\text{ }(k_0\leq -\frac 12E_g)}\psi _e^{+}(k_0,{\bf k)}%
\left| {\bf 0}\right\rangle 
\end{equation}
where $|0>$ is the vacuum state associated with the conduction and valence
bands. Consider next the electron-hole pair orbital state $\varphi _k$ in
which the pair $(k_0,{\bf k};k_0,{\bf -k})$, representing the electron state 
$\left( k_0,{\bf k}\right) $ in the conduction band and its corresponding
hole partner in the valence band, has equal probability of being either
fully or totally empty\cite{b5}:

\begin{equation}
\label{e41}\varphi _k=\frac 1{\sqrt{2}}[1+\psi _h^{+}(k_0,{\bf -k})\psi
_e^{+}(k_0,{\bf k})]|\Phi _0> 
\end{equation}

Here $\psi_e^{+}$ acts over the conduction band states while $\psi_h^{+}$
acts over the valence band states only.

If we consider that all excited electron-hole pair states for all values 
of ${\bf {k}}$ in the two-band semiconductor are arranged in a similar 
way and
are essentially uncorrelated, the state vector $|\Phi >$ for this total
excited state is the linear combination

\begin{equation}
\left| \Phi \right\rangle =\prod\limits_k\varphi _k=\prod\limits_{k(k_0\geq
\frac 12E_g)}\frac 1{\sqrt{2}}[1+\psi _h^{+}(k_0,-{\bf k)}\psi _e^{+}(k_0,%
{\bf k)}]\left| \Phi \right\rangle 
\end{equation}

At finite T, our excited state can therefore be considered a 
superposition of states with different
numbers of occupied electron-hole pairs. This vector state can be 
realized if there is not 
a single electron in the conduction band without its corresponding hole
partner in the valence band. That is, for a given $k_0$, the interband pair 
$(k_0,{\bf k};k_0,{\bf -k})$ has an equal chance of being fully occupied or 
completely empty.

Following of previous work\cite{b5},

\begin{equation}
\label{e43}\psi _e^{+}(k_0,{\bf {k}})|\Phi >=\psi _h(k_0,-{\bf {k}})|\Phi
>\;\;\;, 
\end{equation}
or, equivalently,

\begin{equation}
\label{e44}\psi _e(k_0,{\bf {k}})|\Phi >=\psi _h^{+}(k_0,-{\bf {k}})|\Phi > 
\end{equation}

This means that if we add an extra independent electron to a $k$-state of
the conduction band we automatically exclude the electron-hole state $(k_0, 
{\bf {k}}; k_0, -{\bf {k}})$, across the energy gap, as a possible state for
one existing exciton in the semiconductor. It is thus physically equivalent
to add an electron $(k_0 , {\bf {k})}$ to the conduction band or to remove a
hole from the corresponding state $(k_0, -{\bf {k}})$ in the valence band of
the direct gap semiconductor.

Here as before $\psi_e$ and $\psi_h$ satisfy conventional fermion
anticommutation relations since when $\psi _e$ and $\psi _h$
operate over a direct product of independent vector spaces, which are
associated with the conduction and valence bands of the semiconductor, they
commute with each other.
It follows from this that for the excited state vector $|\Phi>$

\begin{equation}
\label{e45}\{\psi _e(k_0,{\bf {k}}),\psi _h(k_0,-{\bf {k}})\}|\Phi >=\{\psi
_e^{+}(k_0,{\bf {k}}),\psi _h^{+}(k_0,-{\bf {k}})\}|\Phi >=1 
\end{equation}

These unconventional anticommutation relations allow us to
establish the exact supersymmetry of the static part of the effective
hamiltonian density for the elementary excitations in the semiconductor when
it is restricted to act in excited states such as $|\Phi >$ just as we 
did for $T=0$. 

\section{Supersymmetry and Electron-Hole Excitations}

In our derivation of the effective Hamiltonian of fermionic and bosonic 
excitations, the effective potential for bosonic field always take the 
Ginzburg-Landau form at both zero and finite temperature in the weak 
coupling regime when the coupling constant exceeds some 
characteristic critical value. Thus, if we take 
\begin{equation}
\label{e51}V^2(\phi ^{*},\phi )=f^2(\phi ^{*}\phi -R^2)^2 \text{,}
\end{equation}
the static part of Hamiltonian ${\cal {H}_{{\rm st}}}$ of the 
fermion-boson low lying
excitations in the semiconductor consists of the bosonic kinetic 
contribution and its
potential as well as the fermion-boson interaction term expesssed as 

\begin{equation}
\label{e52}{\cal {H}}_{{\rm {st}}}=f^{\prime}\Psi ^{+}\left( \phi \tau _{+}+\phi
^{*}\tau _{-}\right) \Psi ^{+}+V^2\left( \phi ^{*},\phi \right) +\pi ^{*}\pi 
\end{equation}
with $\pi $ and $\pi ^{*}$ being the canonical conjugate to $\phi $ and $%
\phi ^{*}$ respectively and $f^{\prime}=f$ in the case of (2.38) and (3.24).
Clearly, at tree level, the amplitude $\sigma $-mode is such that $m_\sigma
^2=4R^2f^2$ while the $\Theta $-field is the corresponding Nambu-Goldstone
mode with $m_\Theta =0$. In the semiconductor there are therefore two kinds
of bosonic excitations, the radial $\sigma $ oscillations and the {\em %
massless }$\Theta $-phase mode. The fermionic mode is a single-particle
excitation. In the presence of the energy-gap the fermion mass is finite and
equal to $f^{\prime}R$. In contrast, the $\sigma -$mode, which is directly 
associated
with the electron-hole pair, has a mass of $2fR$, while the $\Theta -$mode,
which measures essentially the phase coherence of the electron-hole pair and
which couples to transverse excitations, has zero mass. In the presence of
the Coulomb field the electron-hole pair form an exciton state and the {\em %
massless }$\Theta $- mode is transmuted into a massive plasmon mode. If we
take into consideration the existence of the fermionic mode with mass $fR$ as 
in the case of (2.38) and (3.24),
we establish the mass ratios $0:fR:2fR$ for the fermion-boson low lying
excitations of the semiconductor at both zero and finite temperature. 
This is, of course, the same mass ratio
observed by Nambu in the BCS superconductor \cite{b1,b2}. Hence at 
non-zero temperature, we can still define the fermionic composite charge 
operators $\hat Q$ and $\hat Q^{+}$

\begin{equation}
\ \hat Q^{+}=\pi ^{+}\psi _e-iV\psi _h^{+} 
\end{equation}

\begin{equation}
\hat Q\text{ }=\text{ }\pi \psi _e^{+}+iV\psi _h\text{ } 
\end{equation}
In terms of these operators it then follows that

\begin{equation}
\label{e54}\{\hat Q,\hat Q^{+}\}=\pi ^{+}\pi +V^2+f\phi \psi _e^{+}\psi
_h^{+}+f\phi ^{*}\psi _h\psi _e={\cal H_{{\rm st}}} 
\end{equation}
In order to guarantee the nilpotency of the charge operators we next 
impose a restriction in the 
state vectors they should act. For this reason we limit the action of the 
operators $\hat Q$ and $\hat Q^{+}$ to the symmetric electron-hole excited 
states such as $|\Phi >$.
Invoking the anticommutation relations $\ref{e45}$ we can establish an exact
supersymmetry and a non-trivial mapping to $SUSY$ quantum mechanics.
Following our previous treatment of this problem
one possible representation for the composite operators is then

\begin{equation}
\hat Q=(\pi -iV)\tau _{+} 
\end{equation}

\begin{equation}
\hat Q^{+}=(\pi -iV)\tau _{-} 
\end{equation}
with $\pi =-i\partial /\partial \rho $. In this way the static part of the
effective hamiltonian ${\cal {H}_{{\rm st}}}$ reduces to

\begin{equation}
\label{e55}{\cal {H}}_{{\rm st}}=\left\{ \hat Q,\hat Q^{+}\right\} =\pi
^2+V^2(\rho )+\frac{\partial V(\rho )}{\partial \rho }\tau _3\;\;, 
\end{equation}
where $\partial V/\partial \rho =2f\rho $.

In this form ${\cal {H}_{{\rm st}}}$ is exactly supersymmetric and
is equivalent to $1+1$ $SUSY$ quantum mechanics. The operators $\hat Q$ 
and $\hat Q^{+}$ are now nilpotent and the fermionic and bosonic components 
of the
model have both, at tree level, an equal mass of $2fR$ at both $T=0$ and 
non-zero $T$.

\section{Conclusion}

Using the framework of Nambu for the BCS model of 
superconductivity, we generalized our previous work to take explicit 
account of finite temperature effects in our supersymmetric description 
of elementary excitations in a semiconductor. As expected, despite the 
presence of a non-zero energy gap, if we couple fermions and bosons 
directly to heat bath, supersymmytry is spontaneously broken. This is 
simply due to the fact that in this case we can no longer maintain the 
special mass relation $0:1:2$ between the existent fermionic and bosonic 
elementary excitations in the physical system. However, we argue that the 
electron-hole pair state, as the Cooper pair in a superconductor, should 
not be considered as a real boson having an infinite number of energy states 
simply 
due to the presence of a heat bath. The excited state which is closer in 
energy to the uncorrelated pair state is precisely the one in which at 
least one pair breaking is allowed. Therefore the higher energy bosonic 
state are not physically meaningful for the effective hamiltonian of the 
elementary excitations in the semiconductor.These states are  
automatically ruled out if 
we assume that both fermion and boson excitations are constituted by the 
fermionic single-particles states which couple directly to the heat 
bath. The excitations will therefore feel the temperature effects 
only through 
the new temperature dependent mass and coupling parameters and 
supersymmetry can be preserved if $\lambda$ exceeds a critical value. The 
mapping to $1+1$ $SUSY$ quantum mechanics is then shown to follow the same 
$T=0$ scheme shown in \cite{b5} and the mass parameter ratios are 
essentially preserved at finite temperature.

\section*{Acknowledgments}

This work was financially supported by the Conselho Nacional de
Desenvolvimento Cient\'ifico e Tecnol\'ogico - CNPq and by the Financiadora
de Estudos e Projetos - FINEP. One of us, X.X, was also supported by 
an ICTP(Trieste, Italy) fellowship, and A.F. acknowledges both the 
financial support of the Ministry of Education, Science and Culture 
of Japan and the hospitality of the staff of the Yukawa Institute for 
Theoretical Physics, where part of this work was concluded.

\newpage

\section*{Figure Captions}

\begin{description}
\item[Fig.1 -]  Diagram which generates the boson field free effective
lagrangian.

\item[Fig.2 -]  Diagram which defines the boson field self-interaction.

\item[Fig.3 -]  Diagram which defines the one-loop correction for four-fermion 
coupling constant.
\end{description}

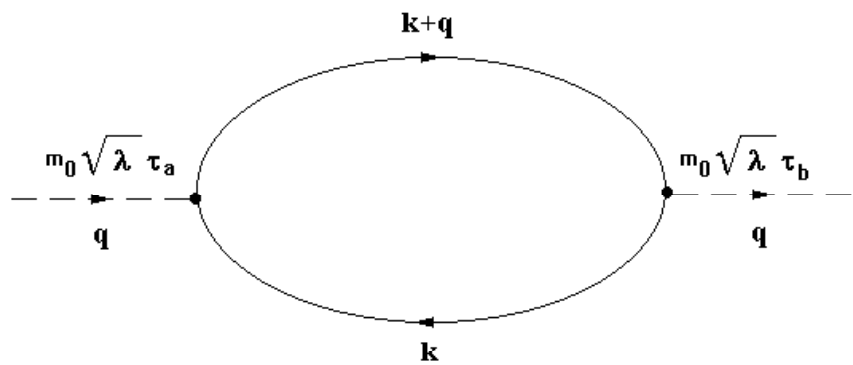
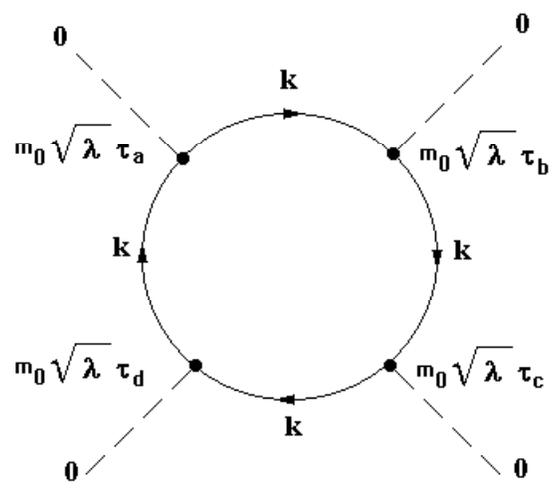
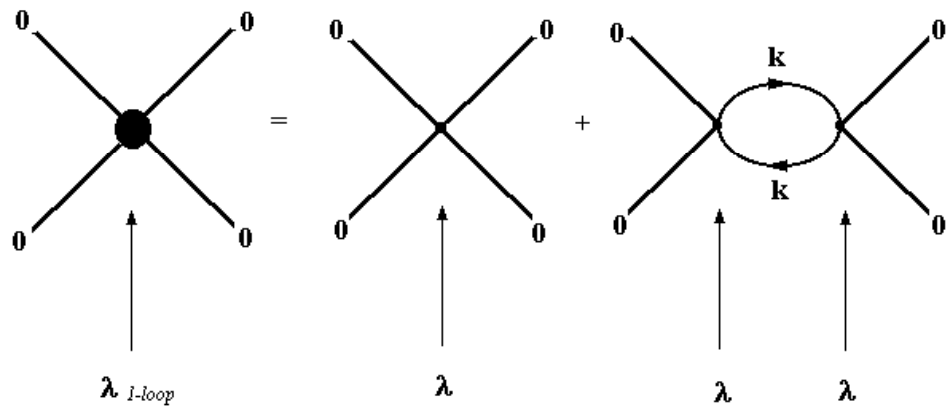
\end{document}